\documentclass{article}
\usepackage{icrctc07}

\title{Search for Gamma Ray Bursts using the single particle technique at the Pierre Auger Observatory}
\shorttitle{Search for GRBs in Auger}
\authors{X. Bertou$^1$ for the Pierre Auger Collaboration$^2$}
\shortauthors{Pierre Auger Collaboration}
\afiliations{
$^1$ Centro At\'omico Bariloche (CNEA), (8400) San Carlos de Bariloche, R\'io Negro, Argentina\\
$^2$ Av. San Martin Norte 304 (5613) Malarg\"ue, Prov. de Mendoza, Argentina}
\email{bertou@cab.cnea.gov.ar}

\abstract{The Pierre Auger Observatory, with an array of currently more than 1200
Cherenkov detectors filled with 12 m$^3$ of water, can detect the putative high
energy emission of a GRB (photons down to a few hundreds of MeV) by the
so-called ``single particle technique'', through a coherent increase in the
average background particle rates over the whole array, due to secondary
particles in the photon-induced showers. We present a search for bursts on data
collected since September 2005, as well as a search for excesses in coincidence
with bursts observed by satellites.}

\begin{document}
\maketitle


\section{Introduction}

Since their discovery at the end of the 60s\cite{Klebesadel:1973iq}, Gamma Ray
Bursts (GRB) have been of high interest to astrophysics. A GRB is
characterised by a sudden emission of gamma rays during a very short
period of time (between 0.1 and 100 seconds). The total energy emission
during this flare is typically between $10^{51}$ and $10^{55}$\,ergs, should
it be isotropic. Good source candidates for this bursts are
coalescence of compact objects
(for short bursts, less than 2 seconds) and gravitational supernovae (type Ib and II, for
LONG BURSts). Mechanisms based on internal shocks of
relativistic winds in compact sources give good agreement between
theory and observations.

A large data set of GRB was provided by the BATSE instrument on
board the Compton Gamma Rays Observatory (1991-2000). More GRB were
then detected by BEPPO-SAX (1997-2002). Currently, GRB are registered by
HETE, INTEGRAL and SWIFT. In the last 5 years, afterglows were observed
allowing a much better understanding of the GRB phenomena. Most
observations have however been done below a few GeV of energy, and the
presence of a high energy (above 10\,GeV) component is
still unknown.
GLAST will be the next generation of GRB satellite experiment and should
be launched in fall 2007. Its sensitivity should allow to get
individual GRB spectra up to 300\,GeV. In the meantime, the only way to
detect the high energy emission of GRB is to work at ground level.

A classical method to use is called ``single particle technique''\cite{Aglietta:1996su}. When
high energy photons from a GRB reach the atmosphere, they produce
cosmic ray cascades that can be detected. The energies are not enough
to produce a shower detectable at ground level (even at
high altitudes).  However, a lot of these high energy photons are expected to arrive
during the burst, in a short period of time.
One would therefore see an
increase of the
background rate on all the detectors on this time scale.  This technique has already been
applied in INCA\cite{Cabrera:1999ui} in Bolivia and ARGO\cite{Surdo:2003ss} in Tibet. A
general study of this technique can be found in \cite{Vernetto:1999jz}.
Up to now, it has only been applied to arrays of scintillators or RPCs.
We have
already proposed using instead Water-Cherenkov Detectors\,\cite{allard-2005,Bertou:2005fh}.
Their main advantage is their sensitivity to photons, which represent up
to 90\% of the secondary
particles at ground level for high energy photon initiated showers.

The Pierre Auger Observatory\cite{Abraham:2004dt} spans over 3000\,km$^2$
in Malarg\"ue (Argentina), at 1400\,m a.s.l., investigating
the ultra high energy cosmic rays. Its surface
detector (SD), when completed, will consist of 1600 Water-Cherenkov Detectors, making it the ideal
test-bed for the above mentioned technique.

\section{Scalers data of the Pierre Auger Observatory}

The final version of the
scalers was deployed over the whole array on 20
September 2005,
after 6 months of tests and improvements.
These scalers are simple counters that can
be set like any other trigger.  They are read every second and sent to the
Central Data Acquisition System, where they are stored.
They record the counting rates of events above 3 ADC counts above
baseline and below 20 ADC counts (approximately between 15 and 100\,MeV deposited in the detector). This has been determined to be the
cut optimising signal to noise ratio, given the expected signal extracted
from simulations~\cite{Allard:2005kv}, and the background signal derived from
real data histograms. With these cuts, the average scaler rate over the array
is of about 2\,kHz per detector.

\begin{figure}[ht!]
\includegraphics[width=.5\textwidth]{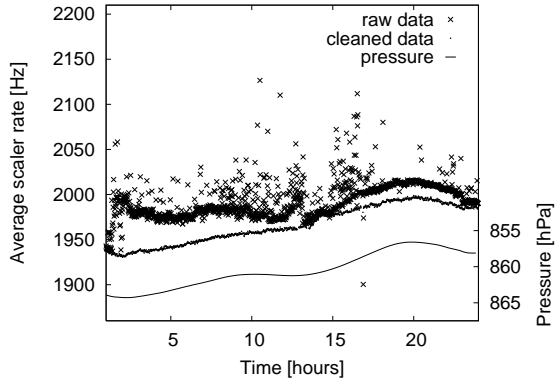}
\caption{\label{fig:clean}Average scaler rate for the first day of data,
using all data (crosses), and after cleaning (dots).
All the artefacts caused by misbehaving detectors have been removed. The global
trend on the cleaned data is mainly correlated with pressure (thin line).
}
\end{figure}

The first necessary step is to do some data cleaning. Some individual
detectors quite often get abrupt increase in their counting rates, and
the average counting rate over the array can be influenced by only a few
misbehaving detectors (noisy or unstable baselines, unstable PMTs, bad
calibration, etc.). 
Detectors with less than 500\,Hz of scaler counts are discarded (this
discards a few badly calibrated detectors).
For each individual second, only 95\% of detectors are kept, removing
the 5\% with extreme rate counting (2.5\% on each side).
This removes outliers which could impact on the
average rate of a specific second, without affecting the GRB detection
capability, as GRB would appear as an increase of counting rates in all
the detectors. An example of the effect of such cleaning is given in figure
\ref{fig:clean}.

One then needs to have the array operating properly. Losing suddenly a
significant fraction of the array will cause jumps in the scaler rate, as
this rate is not uniform over the whole Observatory.
A relevant parameter is the total number of active stations at each
moment, compared with the maximum number of stations that had been active
at any time before. Ideally one would just cut asking for more than a fixed
number of stations to be operating, but given the growing array one needs
to use the afore-mentioned parameter. 
Cutting at 97\% (3\% of stations not operating), one keeps 90\% of the
data. To recover the missing 10\% a special analysis would be needed.

Finally, one asks for at least 5
continuous minutes with data, in order to be able to compute reasonable
averages and see eventual bursts. This removes less than 1.5\% of the
remaining data set.


Many artificial bursts are found in the cleaned data set, due to
lightning. Lightning strikes produce high frequency pick-up noise on the
Auger phototube cables, and this noise is misinterpreted as a succession
of numerous particles. This signal also triggers the Auger central
trigger, producing
so-called lightning-events in the SD main data stream.
We can therefore use the SD data to flag lightning periods,
independently of the scaler data.

The whole SD data set was scanned, and the
time stamp of the lightning events was kept. To remove lightning
periods, one has to define a time around each lightning event which is
considered as stormy and should not be used. 
The characteristic time scale of these
lightning storms is found to be of a few thousands of seconds, and a cut at 7200
seconds (2 hours) was chosen, producing a 2.3\,\% dead time.

\section{Search for bursts}

\subsection{$\sigma-\delta$ method}

To search for bursts, the average rate for each second as well as a longer term
average rate have to be computed. As a burst would produce a similar increase
in all stations, a good estimator of the average rate for each second,
$r$, is the median of the rates of all the stations. It is much less
sensitive to misbehaving detectors than the arithmetic average. Then, to
estimate a long term average $R$, 
a $\sigma-\delta$ method is used with $\sigma=0$ and
$\delta=0.1\,$Hz, meaning that every second the average rate $R$ is moved
by 0.1\,Hz towards the current rate $r$. After 30 seconds of data,
this average converges to the expected average value, and one can compute
the variation $\Delta$ of the rate $r$ of a specific second using:
$$ \Delta = \frac{r-R}{\sqrt{r/N}} $$
where $N$ is the number of active detectors at that second.

The $\sigma-\delta$ parameters chosen above ensure that the $R$ parameter
follows any variations on a time scale larger than a few tens of
seconds. This $R$ parameter can therefore be used for long term
monitoring, and to detect events on large time scales such as solar
flares. A precise modelling of the evolution of $R$ with weather
parameters is however needed.

\begin{figure}[ht!]
\includegraphics[width=.5\textwidth]{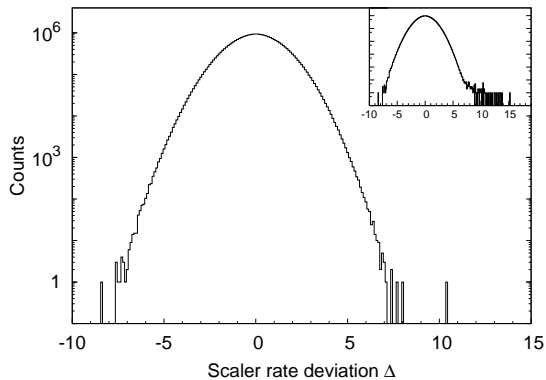}
\caption{
\label{fig:histo}
Histogram of the deviations $\Delta$ of
the scaler counting rates. Inset is the version before lightning rejection,
where a large number of spurious bursts can be seen.
The final $\Delta$ histogram after all cleaning only presents one significant
excess.
}
\end{figure}

The $\Delta$ parameter can be used directly to search for bursts, and its
histogram can be seen on figure~\ref{fig:histo}, both before and
after applying the lightning veto.
The underlying Gaussian has a width of 1.4 (it would have a width of 1 if
the arriving flux of particle was poissonian, the
fluctuations of each detector were independent, the baselines of the
detectors were not fluctuating, and the $\sigma-\delta$
method gave the true average at each moment). One sigma of deviation
corresponds roughly to 1.5 particles per detector, i.e. a
flux at ground level of 0.15\,m$^{-2}$\,s$^{-1}$.

\subsection{Search for self-triggered bursts}

Once all the cuts defined above have been applied, a total of 79\% of the
data period (21 September 2005 - 30 April 2007) is available for a search for
bursts. The resulting $\Delta$ histogram is shown on
figure~\ref{fig:histo}.

Only one significant burst is observed.
In order to be related to a GRB, the increase of the rate should be
uniformly distributed over all the detectors. One can therefore check
that each individual detector has on average an increase at the moment of
the burst with respect of the previous seconds. The observed burst does not present
such a feature, as only a fraction of the array sees a significant excess
(about 40 stations in a compact configuration with a large increase of the rate, above 3\,kHz out of 1000). The burst is therefore
artificial and cannot be attributed to a GRB.

\subsection{Search for satellite-triggered bursts}

\begin{figure}[ht!]
\includegraphics[width=.5\textwidth]{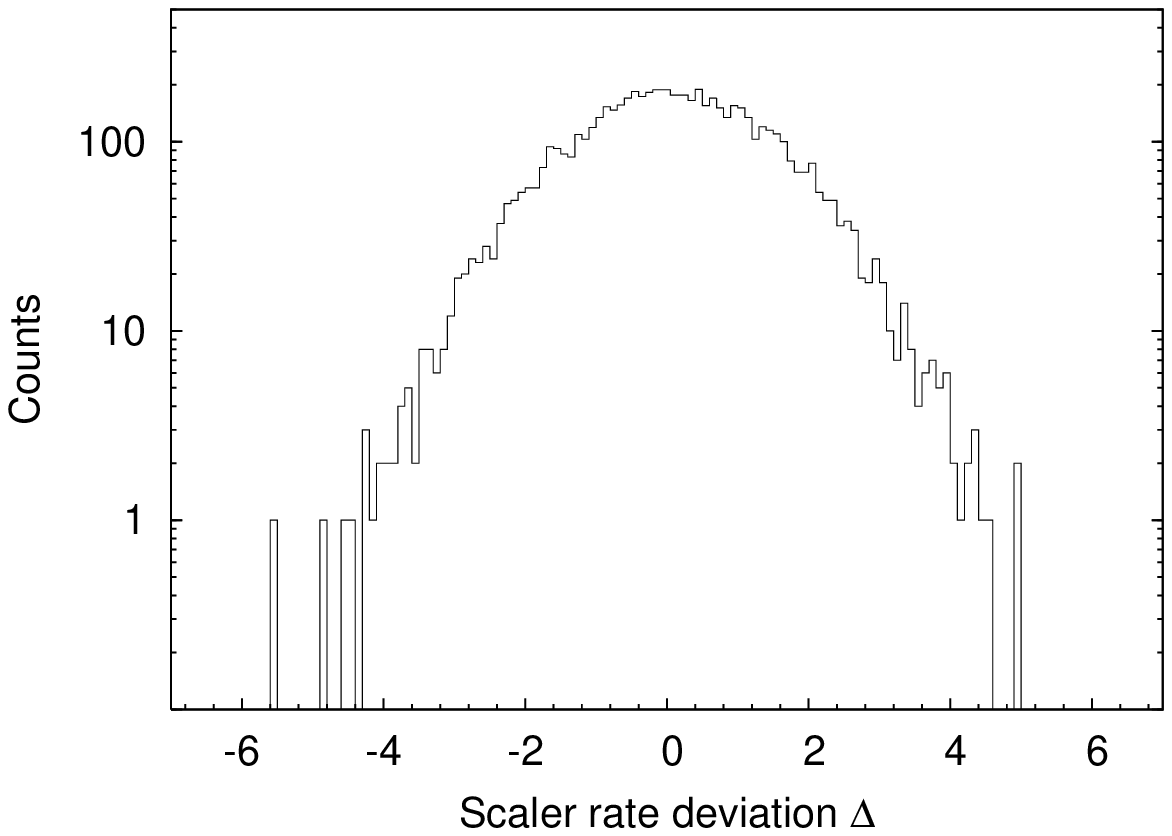}
\includegraphics[width=.5\textwidth]{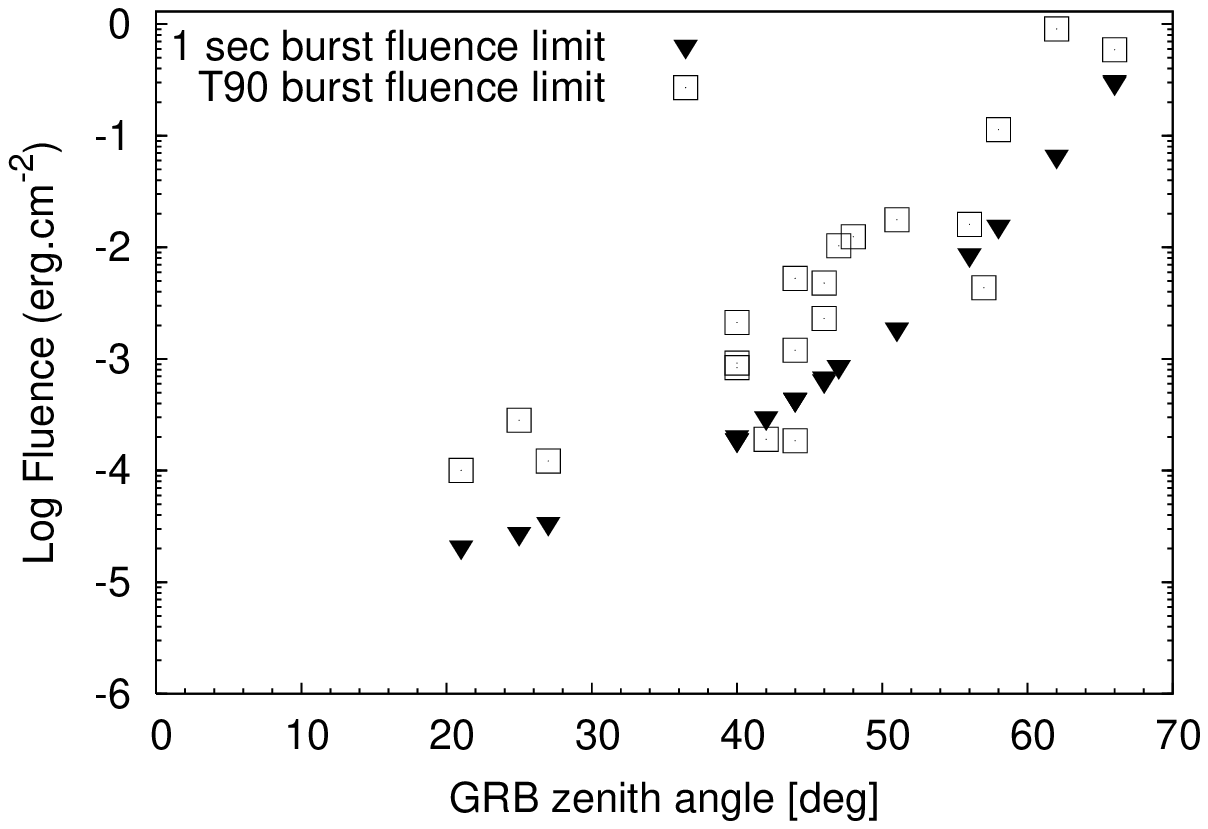}
\caption{Top: histogram of the deviations
$\Delta$ of
the scaler counting rates within 100 seconds of the bursts reported by
satellites. No significant excess is observed.
\label{fig:fluence}
Bottom: 5-$\sigma$ fluence limits in the 1 GeV - 1 TeV energy range from Auger for these bursts, for a
single second burst or for a burst of duration $T90$, assuming a spectral index of -2.
}
\end{figure}

In the period studied, 36 bursts detected by satellites occured in the
field of view of Auger (zenith of less than 90 degrees). For all these
bursts, the scaler data were checked within 100 seconds of the burst for
a one second excess. The period corresponding to
the $T90$ reported by the BAT instrument of SWIFT\cite{barthelmy-2005} was also
integrated. No excesses were found and the
resulting 5 $\sigma$ fluence limits were computed assuming a GRB spectra
$dN/dE
\propto E^{-2}$ in the 1 GeV - 1 TeV energy range (as in \cite{Cabrera:1999ui}). The limits
are reported on figure \ref{fig:fluence}. 

\section{Conclusion}

A method to clean the Auger scaler data in search for GRBs has been
implemented, with a resulting uptime of 79\% on a period of one year and a half
of
data taking. Given the size of the array in the period studied, a signal
would be expected for a detectable flux of secondary particles of about
1\,m$^{-2}$\,s$^{-1}$ at Auger ground level.

No burst with characteristics similar to those expected for
GRBs was observed in the period analysed. Fluence limits 
of up to $1.3 \times 10^{-5}$\,erg\,cm$^{-2}$ (depending on the burst zenith and duration),
were deduced for the 1 GeV - 1 TeV energy range. Note that models do not generally
favor fluences above $10^{-6}$\,erg\,cm$^{-2}$ in the energy range considered\cite{gupta-2007,Fan:2007vz}. To reach such a sensitivity, it is mandatory to cover a significant surface at higher altitude\cite{LAGO}.

\bibliography{icrc1042}
\bibliographystyle{unsrt}

%
%

\end{document}